\documentclass{PoS}

\usepackage{graphicx}
\usepackage{enumerate} 
\usepackage{amsmath,amssymb}

\title{ 
Calculation of the nucleon sigma term and strange quark content
with two flavors of dynamical overlap fermions}

\ShortTitle{
Nucleon sigma term and strange quark content
with two flavors of dynamical overlap fermions }

\author{\speaker{H. Ohki}$^{a,b}$, H. Fukaya$^c$, 
	S. Hashimoto$^{d,e}$, 
	H. Matsufuru$^d$, J. Noaki$^d$, %
        T. Onogi$^b$,  \quad \quad  \quad  \quad  
	E. Shintani$^d$, N. Yamada$^{d,e}$ 
        (for JLQCD collaboration) \\
	\llap{$^a$}Department of Physics, 
         Kyoto University, Kyoto 606-8501, Japan,\\
     	\llap{$^b$} Yukawa Institute for Theoretical Physics, 
	Kyoto University, Kyoto 606-8502, Japan, \\
        \llap{$^c$}The Niels Bohr Institute, 
  	The Niels Bohr International Academy 
   	Blegdamsvej 17 DK-2100 Copenhagen, Denmark, \\
	\llap{$^d$}
	High Energy Accelerator Research Organization (KEK), 
     	 Tsukuba 305-0801, Japan, \\
	\llap{$^e$}
	School of High Energy Accelerator Science, 
  	The Graduate University for Advanced Studies (Sokendai), 
	Tsukuba 305-0801, Japan, \\
	E-mail: \email{ohki@yukawa.kyoto-u.ac.jp}}

\abstract{
 We present a calculation of the nucleon sigma term
 on two-flavor QCD configurations
 with dynamical overlap fermions.
 We analyse the lattice data
 for the nucleon mass using the baryon
 chiral perturbation theory.
 Using partially quenched data sets, we extract
 the connected and disconnected contributions
 to the nucleon sigma term separately.
 Chiral symmetry on the lattice simplifies
 the determination of
 the disconnected contribution.
 We find that the strange quark content,
 which determines the neutralino
 dark matter reaction rate with nucleon
 through the Higgs boson exchange,
 is much smaller than the previous lattice
 results.}

\FullConference{The XXVI International Symposium on Lattice Field Theory\\
		 July 14-19 2008\\
		 Williamsburg, Virginia, USA}

\begin{document}
\section{Introduction}	
Nucleon sigma term $\sigma_{\pi N}$ 
is given by a scalar form factor of nucleon at zero recoil. 
While up and down quarks contribute to $\sigma_{\pi N}$ both as
valence and sea quarks, 
strange quark appears only as a sea quark
contribution. 
As a measure of the strange quark content of the nucleon, 
the $y$ parameter is commonly introduced.
These parameters are defined as 
\begin{equation}  \label{eq:piNsigma}
  \sigma_{\pi N} =  m_{ud} 
  \langle N | \bar{u}u+\bar{d}d | N \rangle, 
\quad \quad 
y  \equiv  
  \frac{2\langle N | \bar{s}s  | N \rangle}
  {\langle N | \bar{u}u +\bar{d}d | N \rangle}.
\end{equation}
The $y$ parameter plays an important role 
to determine the detection rate
of possible neutralino dark matter
 in the supersymmetric extension 
of the Standard Model  
\cite{Griest:1988yr,Bottino:1999ei,
  Ellis:2003cw,Baltz:2006fm,Ellis:2008hf}.
Already with the present direct dark matter search experiments one 
may probe a part of the MSSM model parameter space, 
and new experiments
such as XMASS and SuperCDMS will be able to improve the sensitivity 
by 2--3 orders of magnitude. 
Therefore, a precise calculation of the $y$ parameter 
will be important for excluding or proving 
the neutralino dark matter scenario. 

Using lattice QCD, one can calculate the nucleon sigma term 
directly.
Furthermore, it is possible to determine the valence and sea quark
contributions separately.
Previous lattice results were
$\sigma_{\pi N}$ = 40--60~MeV, $y$ = 0.66(15) \cite{Fukugita:1995ba},
and
$\sigma_{\pi N}$ = 50(3)~MeV, $y$ = 0.36(3) \cite{Dong:1996ec} 
within the quenched approximation, 
while a two-flavor QCD calculation \cite{Gusken:1998wy} gave
$\sigma_{\pi N}$ = 18(5)~MeV and $y$ = 0.59(13).
There is an apparent puzzle in these results: 
the strange quark content 
is unnaturally large compared to 
the up and down contributions 
that contain the connected diagrams too.
 
Concerning this problem, it was pointed out 
that using the Wilson-type fermions, 
the sea quark mass dependence of the additive mass renormalization 
and lattice spacing can give rise to 
a significant lattice artifacts
in the sea quark content~\cite{Michael:2001bv}.
Unfortunately, after subtracting this contamination 
the unquenched result has large statistical error, $y=-0.28(33)$.

In this study, 
we analyze the data of the nucleon mass obtained 
from a two-flavor QCD simulation 
employing the overlap fermion~\cite{Aoki:2008tq} 
which can remove this problem by explicitly maintaining 
exact chiral symmetry on the lattice. 
Although the two-flavor QCD calculation cannot avoid the
systematic error due to the neglected strange sea quarks,
our study with exact chiral symmetry reveals
the underlying systematic effects in the calculation of the nucleon
sigma term, especially in the extraction of its sea quark
contribution.
It therefore provides a realistic test case,
which will be followed by 2+1-flavor calculations in the near
future
\footnote{
  For a very recent result from 2+1-flavor QCD, see
  \cite{WalkerLoud:2008bp}.
}.
The full details of this work is presented in~\cite{Ohki:2008ff}.

\section{Lattice simulation}
\label{sec:Simulation}

We make an analysis of the nucleon mass 
obtained by two-flavor QCD configurations generated with
dynamical overlap fermions.
Our simulations are performed at a lattice spacing $a$ = 0.118(2)~fm
on a $16^3\times 32$ lattice
with a trivial topological sector $Q=0$.
For each sea quark mass, we accumulate 10,000
trajectories; the calculation of the nucleon mass is done at every
20 trajectories, thus we have 500 samples for each
$m_{\mathrm{sea}}$.  
For the sea quark mass $am_{\mathrm{sea}}$ we take six values: 0.015,
0.025, 0.035, 0.050, 0.070, and 0.100 that cover the mass range
$m_s/6$--$m_s$ with $m_s$ the physical strange quark mass.
Analysis of the pion mass and decay constant on this data set 
is found in \cite{Noaki:2008iy}.
In order to improve the statistical accuracy, 
we use the low-mode preconditioning technique. 
The two-point function made of low-lying modes of the overlap-Dirac
operator is averaged over different time slices with 
50 chiral pairs of the low modes.
For the quark propagator, 
we take a smeared source defined by a function 
$\phi(|\vec{x}|)\propto\exp(-A|\vec{x}|)$ with a fixed $A$ = 0.40.
We then calculate the smeared-local two-point correlator
and fit the data with a single exponential function after averaging
over forward and backward propagating states in time.
we take the valence quark
masses $am_{\mathrm{val}}$ = 
0.015, 0.025, 0.035, 0.050, 0.060, 0.070, 0.080, 0.090, and 0.100.

The matrix element defining the nucleon sigma term 
can be related to the quark mass dependence of the nucleon mass using
the Feynman-Hellman theorem, 
which derives the relations between the nucleon mass and 
the quark contents of the nucleon as
\begin{equation}
  \label{eq:deriv_mN}
  \frac{\partial M_N}{\partial m_{\rm val}} 
  = \langle N | (\bar{u}u+\bar{d}d) | N\rangle_{\rm conn},
  \quad \quad  
  \frac{\partial M_N}{\partial m_{\rm sea}} \
  = \langle N | (\bar{u}u+\bar{d}d) | N\rangle_{\rm disc}.
\end{equation}
The subscripts ``conn'' and ``disc'' on the expectation values
indicate that
only the connected or disconnected quark line contractions are
evaluated, respectively. 
In the present study we exploit this indirect method 
to extract the matrix elements 
corresponding to the nucleon sigma term.

Another possible method is to directly calculate the
matrix element from three-point functions with 
an insertion of the scalar density operator 
$(\bar{u}u+\bar{d}d)(x)$.
In principle, it gives a mathematically equivalent result
to the indirect method including lattice artifacts.
Numerical differences could arise only
in the statistical error and the systematic uncertainties 
of the fit ansatz. 

\section{Analysis of the unitary points with baryon chiral 
perturbation theory}
\label{sec:unitary}

\begin{figure}[tbp]
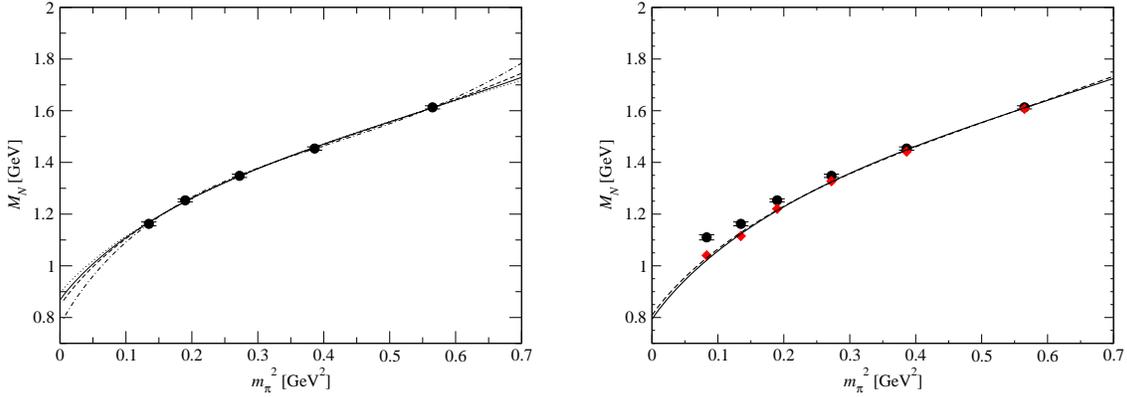

  \centering
  \rotatebox{0}{
    \includegraphics[width=7cm,clip]{./figure/unitary_5pt.eps}
\quad \quad 
    \includegraphics[width=7cm,clip]{./figure/FSE_FIT0.eps}
  }
  \caption{
Chiral fit of the lattice raw data (left)
and the corrected data (right).
In the left panel, the solid curve represents 
the Fit a. The dot, dashed, dot-dashed curves 
correspond to the fits using BChPT including higher order terms.
In the right panel, 
Solid and dashed curves represent the fits using
5 and 6 heaviest data points (diamond), respectively.
For a reference, we also show the raw data (circles).}
\label{fig:fit_FSE_FIT0}
\end{figure}

We carry out baryon chiral perturbation theory (BChPT) 
\cite{Jenkins:1990jv} fits of the nucleon mass 
at the five heaviest quark masses 
for unitary point by using a simplified fit function 
\begin{equation}
  \label{eq:p3}
  M_N = M_0 -4c_1 m_\pi^2 -\frac{3g_A^2}{32\pi f_\pi^2}m_\pi^3 
  +  e_1^r(\mu)  m_\pi^4,
\end{equation}
where $M_0$ is the nucleon mass in the chiral limit and
$f_\pi$ is the pion decay constant fixed 
at its physical value 92.4~MeV and 
the constant $g_A$ describes the nucleon axial-vector coupling.
Since the coupling $g_A$ is very well known experimentally, 
we attempt two options: (Fit a) a fit with fixed $g_A$ (=1.267), and
(Fit b) a fit with $g_A$ being dealt as a free parameter.
Moreover we also carry out the full $\mathcal{O}(p^3)$ fit and 
$\mathcal{O}(p^4)$ fit following the analysis done 
in ~\cite{Procura:2003ig}.
The left panel of Fig.~\ref{fig:fit_FSE_FIT0}
shows the BChPT fits. 
The lattice data show a significant curvature
towards the chiral limit.
All the details of our analysis can be found 
in \cite{Ohki:2008ff}.

In order to estimate the systematic uncertainties 
due to the finite volume effect ,
we correct the data for the finite volume effect using 
BChPT $\mathcal{O}(p^4)$ formula~\cite{Ali Khan:2003cu}.
For the input parameters $M_0$, $g_A$, $c_1$, we use
the nominal values ($M_0$ = 0.87~GeV, $g_A$ = 1.267, 
$c_1=-1.0$ GeV$^{-1}$). 
The chiral fit is then made for the corrected data points using the
formula (\ref{eq:p3}) for the five or six heaviest data points. 
The result is shown in the right panel of
Fig.~\ref{fig:fit_FSE_FIT0}.
After correcting the finite volume effect, 
there are 5-8\% 
decrease in $M_0$ and 4-7\% increase 
in the magnitude of the slope $|c_1|$. 
We observe that the deviation 
due to the finite volume effect is smaller than the
uncertainty of the fit forms.

Taking the Fit a 
($g_A$ fixed, Finite volume corrections not included) as our best fit
and using the variation with fit ansatz and Finite volume corrections
as an estimate of 
the systematic errors,  
our result of the nucleon sigma term is 
\begin{equation}
  \sigma_{\pi N} = 52(2)_{\rm stat}(^{+20}_{-\ 7})_{\rm
    extrap}(^{+5}_{-0})_{\rm FVE}
  \mathrm{~MeV}, 
  \label{eq:result-sigma}
\end{equation}
where the errors are the statistical and the systematic due to 
the chiral extrapolation (extrap) and finite volume effect (FVE).
The largest uncertainty comes from the chiral extrapolation.
The finite volume effect is sub-leading, which is about 9\%.
This result is in good agreement with the phenomenological analysis
based on the experimental data.
%
%
%
\section{Partially quenched analysis and extraction of $y$}
\label{sec:PQChPT}
In order to extract 
the connected and disconnected diagram contributions separately,
we fit the quark mass dependence of the nucleon mass with the
one-loop partially quenched ChPT formula \cite{Chen:2001yi} 
\begin{small}
\begin{eqnarray}
  \label{eq:PQChPT}
  M_N & = &  B_{00} + B_{10}(m_\pi^{vv})^2 +B_{01}(m_\pi^{ss})^2
       +B_{20}(m_\pi^{vv})^4  
       +B_{11}(m_\pi^{vv})^2 (m_\pi^{ss})^2
       +B_{02}(m_\pi^{ss})^4  \nonumber \\
  & &    -\frac{1}{16\pi f_\pi^2}
        \Bigl\{ \frac{g_A^2}{12}
        \left[ -7(m_\pi^{vv})^3 +16 (m_\pi^{vs})^3 
               +9 m_\pi^{vv}(m_\pi^{ss})^2 )\right] 
       + \frac{g_1^2}{12}
       \left[ -19(m_\pi^{vv})^3 +10 (m_\pi^{vs})^3 
              +9 m_\pi^{vv}(m_\pi^{ss})^2 )\right]   \nonumber \\
  & &  + \frac{g_1 g_A}{3}
        \left[ -13(m_\pi^{vv})^3 + 4(m_\pi^{vs})^3 
               +9 m_\pi^{vv}(m_\pi^{ss})^2 )\right]
        \Bigr\} 
\end{eqnarray}
\end{small}%
where $m_\pi^{vv}$, $m_\pi^{vs}$, and $m_\pi^{ss}$ denote the pion
mass made of valence-valence, valence-sea, and sea-sea quark
combinations, respectively.
At this order, one can rewrite the expression
in terms of $m_{\mathrm{val}}$ and $m_{\mathrm{sea}}$ 
using the leading-order relations.
The coupling constant $g_1$ is another 
axial-vector coupling.
We use nominal values 
$g_A=1.267$ and $g_1=-0.66$ 
when they are fixed in the fit.
There are also contributions from the decuplet
baryons. 
In our analysis these contributions can be absorbed 
into the analytic terms in (\ref{eq:PQChPT}). 
The independent fit parameters are
$B_{00}$, $B_{01}$, $B_{10}$, $B_{11}$, $B_{20}$, $B_{02}$, 
$g_1$, and $g_A$. 
We attempt a fit with fixed $g_A$ and $g_1$ (Fit PQ-a), 
a fit with fixed $g_A$ (Fit PQ-b) 
and a fit with all the free parameters (Fit PQ-c). 
Finite volume corrections are not taken into account.

\begin{figure}[tbp]
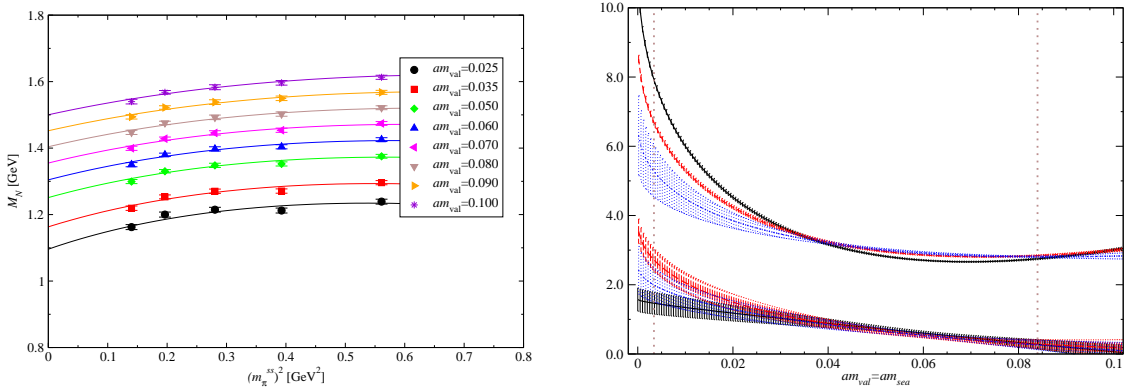

  \centering
  \rotatebox{0}{
    \includegraphics[width=7cm,clip]{./figure/pqch_5-8_b.eps}
\quad \quad
\includegraphics[width=7cm,clip]{./figure/Fig7.eps}
  }
  \caption{
The left panel shows 
partially quenched nucleon masses and fit curves (Fit PQ-b).
The right panel shows 
the connected (up) and disconnected (down) contributions 
to the sigma term evaluated at
$m_{\mathrm{val}}=m_{\mathrm{sea}}$.
Solid, dashed and dotted curves   
represent the results from the Fit PQ-a, PQ-b and PQ-c, 
respectively.   
The two vertical lines show physical up and down quark mass
$am_q=0.0034$ (left) 
and strange quark mass $am_q=0.084$ (right). 
}
\label{fig:conn_and_disc}
\end{figure}

The left panel of Fig.~\ref{fig:conn_and_disc} demonstrates 
the result of the partially quenched ChPT fit.
The sea quark mass dependence 
at eight fixed valence quark masses 
are nicely fitted with the formula (\ref{eq:PQChPT}).
We find that both axial-couplings can be determined with
reasonable accuracy ($g_A=0.93(22)$ and $g_1=-0.29(5)$ )
for the Fit PQ-c.
The right panel of Fig.~\ref{fig:conn_and_disc}
shows the partial derivatives in (\ref{eq:deriv_mN}) 
with respect to $m_{\mathrm{val}}$ and to $m_{\mathrm{sea}}$
evaluated at the unitary points
$m_{\mathrm{val}}=m_{\mathrm{sea}}$.
For both contributions, we clearly find an enhancement towards the
chiral limit.
Results with different fit ansatz show slight disagreement near the
chiral limit, which indicate the size of the systematic uncertainty.
We find that the sea quark content of the nucleon 
$\langle N|(\bar{u}u+\bar{d}d)|N\rangle_{\mathrm{disc}}/
\langle N|(\bar{u}u+\bar{d}d)|N\rangle_{\mathrm{conn}}$
is less than 0.4 for the entire quark mass region in our study, 
so that the valence quark content is 
the dominant contribution to the sigma term.

Rigorously speaking, it is not possible to extract the strange quark
content 
$\langle N | \bar{s}s | N \rangle$ within two-flavor QCD.
Instead, in this work, we provide a ``semi-quenched'' estimate of the
strange quark content which is defined as 
the ratio of the strange quark content 
(disconnected contribution at
$m_{\mathrm{val}}=m_{\mathrm{sea}}=m_s$) to the up and down quark
contributions (connected plus disconnected contributions at
$m_{\mathrm{val}}=m_{\mathrm{sea}}=m_{ud}$) following 
\cite{Michael:2001bv}.
Taking the result from the Fit PQ-b as a best estimate,  
we obtain the parameter $y$ as 
\begin{equation}
  \label{eq:y_result}
  y^{N_f=2} = 0.030(16)_{\rm stat}(^{+6}_{-8})_{\rm
  extrap}(^{+1}_{-2})_{m_s},
\end{equation}
where the errors are statistical and systematic 
from chiral extrapolation and 
from the uncertainty of $m_s$, respectively.  
The chiral extrapolation error 
is estimated by the differences of the results of Fit PQ-a,b and c, 
We also note that there may be an additional $\sim$10\% error from
the finite volume effect as discussed in Section~\ref{sec:unitary}, 
but it is much smaller than 
the statistical error in our calculation.

\section{Discussion and summary}
\label{sec:Discussion}

\begin{figure}
\begin{center}
\begin{tabular}{c}
\rotatebox{0}{
\includegraphics[width=4.cm]{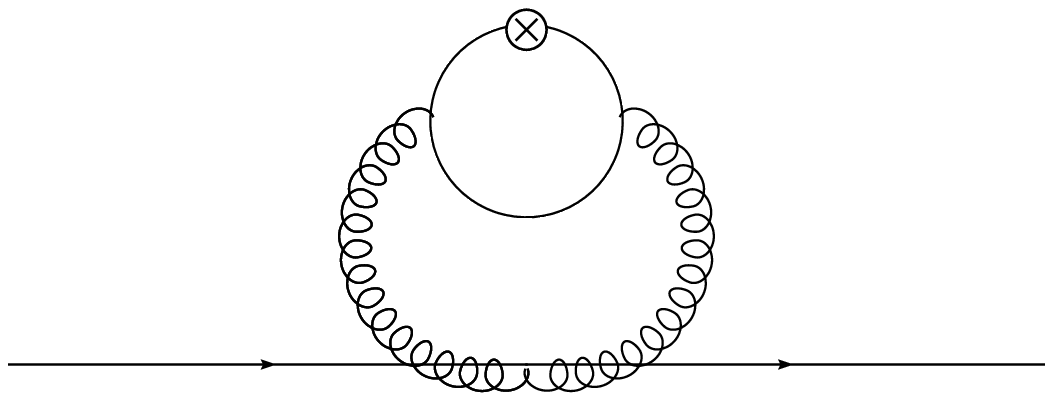}
\quad \quad \quad \quad \quad \quad \quad
\includegraphics[width=4.cm]{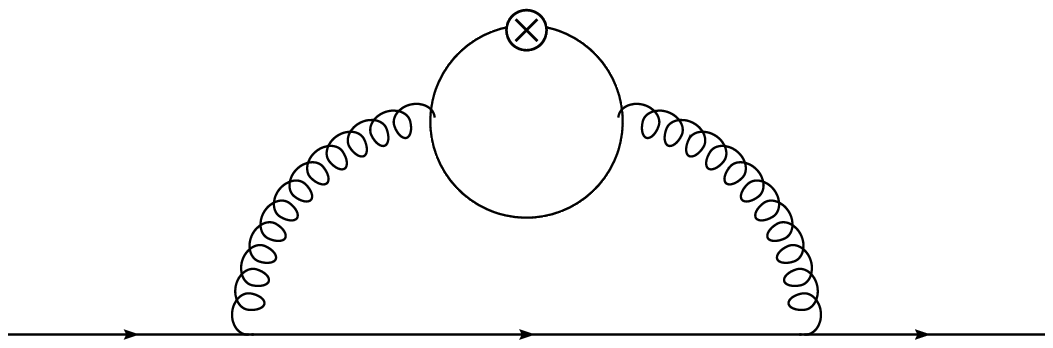}
}
\end{tabular}
\end{center}
\caption{Two-loop diagrams for the mass shift 
and the operator mixing 
involving both the valence and the sea quarks.
These graphs show the corresponding contribution 
to the sigma term. 
Due to the power divergent mass shift, 
the flavor singlet sea
quark scalar operator insertion contributes as 
the flavor singlet scalar operator insertion 
with coefficient of additive mass shift.
} \label{fig:mixing}
\end{figure}

We found that the disconnected contribution 
to the sigma term 
is much smaller than the previous lattice calculations 
with the Wilson-type fermions $y\simeq 0.36\sim 0.66$ 
\cite{Fukugita:1995ba,Dong:1996ec,Gusken:1998wy}
(except for \cite{Michael:2001bv} as explained below).
The authors of \cite{Michael:2001bv} found 
that the naive calculation with the Wilson-type fermions
may over-estimate the sea quark content 
due to the mixing effect from the additive mass shift as 
\begin{equation}
\left.\frac{\partial M_N}{\partial m_{\rm sea}^{\rm bare}}
\right|_{m_{\rm val}^{\rm bare}}
=
Z_m 
\left[
\left. 
\frac{\partial M_N}{\partial m_{\rm sea}^{\rm phys}}
\right|_{m_{\rm val}^{\rm phys}}
+
\left.
\frac{\partial m_{\rm val}^{\rm phys}}
{\partial m_{\rm sea}^{\rm bare}}
\right|_{m_{\rm val}^{\rm bare}} \centerdot
\left.
\frac{\partial M_N}{\partial m_{\rm val}^{\rm phys}}
\right|_{m_{\rm sea}^{\rm phys}}
\right], 
\end{equation}
where $Z_m$ is the renormalization factor of the quark mass. 
Therefore, in order to obtain the derivative (\ref{eq:deriv_mN}) one
must subtract the unphysical contribution from the additive mass.
They found that their unsubtracted result $y=0.53(12)$ is 
substantially
reduced and becomes consistent with zero: $y=-0.28(33)$.
It should be noted that this mixing effect is not limited  
to the spectrum method but arises also in the direct matrix element 
method. It is easy to see that the operator is indeed 
a derivative of the mass shift with respect to 
the sea quark mass so that 
\begin{equation}
\left( \bar{\psi} \psi \right)_{\rm sea}^{\rm bare} 
= 
Z_m 
\left[
\left( \bar{\psi} \psi \right)_{\rm sea}^{\rm phys}
+
\left.\frac{\partial m_{\rm val}^{\rm phys}}
{\partial m_{\rm sea}^{\rm bare}}
\right|_{m_{\rm val}^{\rm bare}}
\left( \bar{\psi} \psi \right)_{\rm val}^{\rm phys}
\right]
\end{equation}
holds. 
(Subtraction of the divergent counter terms is
assumed on the left hand side.)  
Fig.~\ref{fig:mixing} shows
the two-loop contribution of the operator mixing 
of the bare sea quark operator 
with physical valence quark operator
due to the additive mass shift 
which arises from the absence of the chiral symmetry.
Our calculation using the overlap fermion is free from this 
artifact because of the exact chiral symmetry.
Therefore, the small value of $y$ obtained in our analysis
(\ref{eq:y_result}) provides a much more reliable estimate than the
previous lattice calculations
\footnote{Strictly speaking, there is another artifact from 
the sea quark mass dependence of the lattice spacing. 
We have found that the lattice spacing dependence is negligibly 
small, since we exploit mass-independent renormalization scheme
(c.f. \cite{Shintani:2008qe}) in contrast to
\cite{Michael:2001bv}, in which they used the Sommer scale 
for the scale input at each sea quark mass. }.

In summary, 
we have calculated the nucleon sigma term and 
the strange quark content of nucleon 
in two-flavor QCD simulation on the lattice 
with exact chiral symmetry.
Owing to the exact chiral symmetry, our lattice calculation is free
from the large lattice artifacts coming from the additive mass shift
present in the Wilson-type fermion formulations.
From an analysis of partially quenched lattice data,
 we have found that the
sea quark content of the nucleon is less than 0.4 
for the entire quark
mass region in our study.

An obvious extension of this work is the calculation including
the strange quark loop in the vacuum.
Simulations with two light and one strange dynamical overlap quarks
are on-going \cite{Hashimoto:2007vv}.

\vspace{2mm}
The main numerical calculations were performed on 
IBM System Blue Gene Solution 
at High Energy Accelerator Organization (KEK) under support
of its Large Scale Simulation Program (No.~07-16).
We also used NEC SX-8 at Yukawa Institute for Theoretical Physics
(YITP), Kyoto University and at Research Center for Nuclear Physics
(RCNP), Osaka University. 
The simulation also owes to a gigabit network SINET3 supported by
National Institute of Informatics for efficient data transfer through
Japan Lattice Data Grid (JLDG).
This work is supported in part by the Grant-in-Aid of the
Ministry of Education (Nos.  18340075, 18740167,  
19540286,  19740121, 19740160,
20025010, 20039005,  20740156).
The work of HF is supported by Nishina Memorial Foundation.

\end{document}